\documentstyle[aps,epsf,eqsecnum,multicol]{revtex}

\begin{document}
\newcommand{\hgo}{$^{190}$Hg}
\newcommand{\hgt}{$^{192}$Hg}
\newcommand{\hgf}{$^{194}$Hg}
\newcommand{\Pbt}{$^{192}$Pb}
\newcommand{\Pbf}{$^{194}$Pb}
\newcommand{\Pbs}{$^{196}$Pb}
\newcommand{\Pbe}{$^{198}$Pb}
\newcommand{\J}{{\cal J}^{(2)}}
\newcommand{\w}{\omega_{\rm rot}}
\newcommand{\ep}{\epsilon}
\newcommand{\tqp}{{\sc 2qp}}
\newcommand{\bra}[1]{\langle #1 |}
\newcommand{\ket}[1]{| #1 \rangle}

\draft

\title{Microscopic Structure of High-Spin Vibrational States\\
in Superdeformed A=190 Nuclei
}
\author{Takashi~Nakatsukasa
}
\address{AECL, Chalk River Laboratories, Chalk River,
Ontario K0J 1J0, Canada}
\author{Kenichi~Matsuyanagi}
\address{Department of Physics, Kyoto University, Kyoto 606-01,
Japan}
\author{Shoujirou~Mizutori}
\address{
JIHIR, Oak Ridge National Laboratory, P.O.Box 2008, Oak Ridge,
TN 37831, U.S.A.}
\author{Yoshifumi~R.~Shimizu}
\address{
Department of Physics, Kyushu University, Fukuoka 812, Japan}


\maketitle

\tighten
\begin{abstract}
Microscopic RPA calculations based on the cranked shell model are
performed to investigate the quadrupole and octupole correlations
for excited superdeformed (SD) bands in even-even A=190 nuclei.
The $K=2$ octupole vibrations are predicted to be the lowest
excitation modes at zero rotational frequency.
The Coriolis coupling at finite frequency produces different effects
depending on the neutron and proton number of nucleus.
The calculations also indicate
that some collective excitations may produce
$\J$ moments of inertia almost identical to those of the yrast SD band.
An interpretation of the observed excited bands invoking the
octupole vibrations is proposed,
which suggests those octupole vibrations may be prevalent
in even-even SD A=190 nuclei.
\end{abstract}

\bigskip

\begin{multicols}{2}
\section{Introduction}
\label{sec: intro}

Superdeformed (SD) rotational bands provide us
with an opportunity of studying
a finite quantum many-body system at the limits of
a strong Coriolis field and large deformation.
As an example of elementary excitations in rapidly
rotating nuclei,
low-frequency vibrational motion at high spin is of great interest.
Since the large deformation and the rapid rotation of SD bands may
produce a novel shell structure,
we expect features of surface vibrational motions
quite different from those of spherical and normal-deformed (ND)
nuclei.
In this paper quadrupole and octupole vibrations built on the SD
yrast band are discussed in terms of a microscopic
model based on the cranked mean field extended by
the random-phase approximation (RPA).

In a recent paper\cite{NMMS96} (to which we refer hereafter as
NMMS),
we have discussed the quadrupole and octupole correlations
in excited SD bands in $^{190,192,194}$Hg.
We have found that the $K=2$ octupole vibrations are
the lowest excitation modes in these SD nuclei
and the interplay between rotation and vibrations produces
different effects depending on neutron number.
From a comparison with the experimental $\J$ moments of inertia,
we have proposed a new interpretation
that most of the observed excited bands
are the ($K=2$) octupole vibrations.
In this paper, we extend this work to all even-even SD nuclei observed
in A=190 region and compare our theoretical results with new experimental
data.
Assuming that the observed excited SD bands correspond to
the lowest octupole vibrations,
the observed properties (Routhians, $\J$ moments of inertia,
and linking transitions into the yrast SD bands)
will be consistently explained for all even-even A=190 nuclei.

\section{Collective excitations in rapidly rotating
superdeformed nuclei}
\label{sec:general}
Before discussing explicit examples of excited SD bands,
let us discuss the characteristics
of elementary excitation modes
in high-spin SD and ND states in even-even nuclei.

ND nuclei belong to a family of nuclei with ``open-shell''
configurations.
At low spin,
since the pairing correlations produce an energy gap $\Delta$
(typically 1 MeV),
the excitation energies of the lowest two-quasiparticle (\tqp) states
are at least $2\Delta\approx 2$ MeV
and the collective (isoscalar) excitations
become the lowest excitation modes
which have been universally observed in experiments.
However, this situation is changed at high spin by
many rotationally aligned \tqp\ states coming down rapidly
with frequency.
These aligned \tqp\ bands erase the energy gap
and become the dominant modes of low-energy excitation.
Thus, at present,
very few experimental data are available for the
collective modes of excitation at high spin.

On the other hand,
SD nuclei are characterized as ``closed-shell'' nuclei
with a large shell gap near the Fermi surface.
In addition,
the large deformation tends to reduce the aligned angular momenta
of high-$j$ orbitals.
These properties of the SD shape keep
an energy gap in the quasiparticle spectra
even in the high-spin region.
(see quasiparticle Routhians in NMMS and compare them with those of
ND nuclei, e.g., in Ref.\cite{BFM86}).
As a result, the collective states may be still the lowest modes
even at high spin.
In this sense, the SD nuclei could provide an observatory of
the collective excitations in rapidly rotating systems.

In the A=150 region,
the behavior of excited SD bands have been well accounted for
as single-particle
excitations except for a few cases in $^{152}$Dy\cite{Dag95,NMMN95},
$^{148}$Gd\cite{Ang96}, and $^{150}$Gd\cite{Twi96}.
This may be because the pairing correlations are extremely suppressed
in the A=150 region.
Since the pairing correlations may
increase the quadrupole and octupole collectivity,
we expect the different feature of near-yrast excitation spectra
in the A=190 region.

In even-even SD A=190 nuclei,
most of excited bands have $\J$ moments of inertia very similar to those
of the yrast SD bands, which show a gradual increase with frequency.
On the other hand, atypical $\J$ moments
almost constant with frequency
have been observed universally in odd-A SD nuclei
and have been explained by invoking occupation of a quasiparticle state
associated with neutron $N=7$ ($j_{15/2}$)
orbitals\cite{Joy94,Car95,Hug95,Far95}.
These experiments provide important information on the quasiparticle
spectra around the $N=112$ SD shell gap,
which tells us the $N=7$ orbitals is the lowest at high frequency and
the blocking of the $N=7$ orbitals leads to the lack of
alignment producing a constant $\J$.
If we try to interpret the excited bands in even-even nuclei
as the simple \tqp\ states,
it is a puzzle why the similar atypical $\J$
have not been observed in the even-even cases.
We will show in the next section that
the collective excitations can provide a possible answer
to this mystery.

\section{Microscopic structure of collective excitations}
\label{sec:theory}

In this section, we give a brief review of our theoretical model
and discuss the $\J$ moments of inertia for the excited bands.
See NMMS for a complete description of the model
and details of numerical calculations.

In the cranked shell model extended by the RPA theory,
vibrational excitations built on the rotating vacuum are microscopically
described by
superpositions of a large number of \tqp\ states.
The RPA also allows us to describe non-collective
\tqp\ excitations
and weakly collective states which are difficult to discuss
by means of macroscopic models.
Effects of the Coriolis force on these various modes of excitation
are automatically taken into account in RPA solutions
since the mean field is affected by the cranking term
$-\omega_{\rm rot}J_x$.

The model Hamiltonian is assumed to be of the form:
\begin{equation}
 \label{hamiltonian}
 H'= h'_{\rm s.p.} + H_{\rm int}\ ,
\end{equation}

\vspace{-0.2cm}
\noindent
where $h'_{\rm s.p.}$ is a cranked single-particle Nilsson Hamiltonian
including the pairing field.
The quadrupole deformation and pairing gaps are determined by
the standard Strutinsky procedure at $\w =0$.
We have adopted the phenomenological prescription given in
Ref.\cite{WSNJ90} for the pairing gaps at finite frequency
(which is characterized by a ``critical'' frequency $\omega_c$).
The residual interactions are separable multipole interactions,
\begin{equation}
 \label{residual}
 H_{\rm int} = H_{\rm pair} -\frac{1}{2} \sum_{\lambda K}
  \chi_{\lambda K} Q_{\lambda K}^\dagger Q_{\lambda K} \ ,
\end{equation}

\vspace{-0.2cm}
\noindent
where
$H_{\rm pair}$ is a residual pairing interaction consistent with the
mean field and
$Q_{\lambda K}$ are
multipole operators defined in doubly-stretched coordinates
$r_i^{''}=(\omega_i/\omega_0)r_i$ ($i=x,y,z$)\cite{SK89}.
The coupling strengths of the interactions
$\chi_{\lambda K}$ are determined in the same way as in NMMS.
However,
since we do not know exactly the selfconsistent interactions
in a realistic Nilsson potential,
we have done the calculations with two different values of
the octupole coupling strengths;
the ``harmonic-oscillator'' value ($f_3=1$)\cite{SK89} and
the one increased by 5\% ($f_3=1.05$).
We use this symbol $f_3$ as a scaling factor of the coupling strengths.
See eq.(3.13) in NMMS.

After diagonalizing the Hamiltonian (\ref{hamiltonian})
with the RPA theory,
it is written as
\begin{equation}
 \label{rpa-hamiltonian}
 H' = \hbox{const.} +
  \sum_{\alpha,n} \hbar\Omega_n^\alpha X_n^{\alpha\dagger}X_n^\alpha \ ,
\end{equation}

\vspace{-0.2cm}
\noindent
for even-even nuclei at finite rotational frequency.
Here $\alpha$ (=0,1) indicates the signature quantum number.
${X_n^\alpha}^\dagger$ and $\hbar\Omega_n^\alpha$ are the n-th
RPA-normal-mode creation operator and its excitation energy, respectively.
Since we take the yrast SD band as the RPA vacuum,
$\hbar\Omega_n^\alpha$ gives the Routhian (excitation energy in the
rotating frame) relative to the yrast SD band.

From the RPA eigenenergies,
we calculate the $\J$ moments of inertia for excited SD bands
as follows (see NMMS for detail):
The relative difference between the excited and yrast bands
is given by $j^{(2)}=-d^2 \hbar\Omega_n / d \w^2$,
and added to the
experimental $\J_0$ of the yrast band, $\J_0 + j^{(2)}$.
This procedure allows us to take into account the complex correlations
which are implicitly included in $\J_0$; e.g., the pairing fluctuation,
the higher-order pairing.

In the RPA theory, excited states $\ket{n}$ ($n\neq 0$)
are described by superposition
of \tqp\ excitations,
\begin{equation}
 \ket{n} = \sum_{\mu\nu} \left\{
  \psi_n(\mu\nu) a_\mu^\dagger a_\nu^\dagger
 +\varphi_n(\mu\nu) a_\nu a_\mu \right\} \ket{0} \ ,
\end{equation}

\vspace{-0.2cm}
\noindent
where $\ket{0}$ is the RPA vacuum and $\psi_n(\mu\nu)$
($\varphi_n(\mu\nu)$) are the RPA forward (backward) amplitudes.
The backward amplitudes are generally smaller than
the forward amplitudes if the mean field is stable enough.
A non-collective state has a single dominant \tqp\ 
component, namely $|\psi_n(\sigma\rho)|^2 \approx 1$ and
$|\psi_n(\mu\nu)|^2 \approx 0$ for $(\mu\nu)\neq (\sigma\rho)$.
On the other hand, a collective state may be characterized by
the coherent contributions from many different \tqp\ components
and substantial contributions from backward amplitudes.

As is discussed in sec.\ref{sec:general},
the lowest $N=7$ neutron quasiparticles change
$\J$ moments of inertia most significantly.
Therefore, the observed excited bands in even-even nuclei,
which show the $\J$ identical to those of the yrast $\J_0$,
cannot be the lowest \tqp\ states
associated with these orbitals.
A possible explanation for this similarity of $\J$ is
interpreting them as \tqp\ excitations invoking high-$K$ orbitals.
For instance, bands 2 and 3 in \hgf\ have been originally assigned
as two-quasineutron excitations [624 9/2]$\otimes$[512 5/2]\cite{Ril90}.
However, this does not explain why we have not observed \tqp\ 
excitations associated with the neutron $N=7$ orbitals which should
be lower in energy at high frequency (see discussion in
sec.\ref{sec:general}).

Instead, we presume those excited bands could be
the collective states.
First of all, the excitation energies of collective states may be
lower than any \tqp\ state.
In addition,
some collective bands may produce the $j^{(2)}\approx 0$
(which means $\J\approx \J_0$)
because of the spreading of the RPA amplitudes over many \tqp's.
In a highly collective state,
the amplitude $|\psi_n(\mu\nu)|$ of each \tqp\ component
is small and the peculiar effect of the neutron $N=7$
orbitals can be {\it smeared} by the other components.
This ``smearing'' has been actually demonstrated for \hgf\ in NMMS,
and will be also done for Pb isotopes in sec.~\ref{sec:pb}.


\section{Collective excitations in H\lowercase{g} nuclei}
\label{sec: hg}
In this section, results of the RPA calculations are presented
for the excited states in SD Hg nuclei.
Since we have already published this result in NMMS,
here we briefly review the main results of NMMS and
discuss new experimental data on \hgf.
The main conclusions of NMMS are summarized as follows.
\vspace{-0.2cm}
\begin{enumerate}
\item The $K=2$ octupole vibrations are the lowest excitation modes
($E_x \approx 1$ MeV, B($E3;\ 0^+ \rightarrow 3^-) \approx 10$ s.p.u.)
in these Hg isotopes at $\w = 0$.
\vspace{-0.2cm}
\item The $\gamma$ vibrations are higher in energy
and less collective
($E_x \geq 1.4$ MeV, B($E2;\ 0^+ \rightarrow 2^+) < 2\sim 3$ s.p.u.)
than the lowest octupole vibrations at $\w = 0$.
\vspace{-0.2cm}
\item In \hgo, two observed excited SD bands, bands 2 and 4,
are assigned to the lowest octupole
bands with signature $\alpha=1$ and 0, respectively.
The $\alpha=1$ octupole vibration is rotationally aligned,
while the $\alpha=0$ is crossed by a two-quasineutron band
at high frequency.
\vspace{-0.2cm}
\item In \hgt, bands 2 and 3 are assigned to the lowest ($K=2$)
octupole bands with signature $\alpha=1$ and 0, respectively.
Both bands are crossed by a two-quasineutron band at high frequency.
\vspace{-0.2cm}
\item In \hgf, bands 2 and 3 are assigned to the lowest ($K=2$)
octupole bands with signature $\alpha=0$ and 1, respectively.
\vspace{-0.2cm}
\item The strongest mixture of low-$K$ components ($K=0$ and 1)
at finite frequency is predicted for band 2 of \hgo,
which explains why strong dipole decay into the yrast SD band
has been observed only for this band.
\end{enumerate}

For \hgf, recent {\sc gammasphere} experiments have revealed
the excitation energies and spins
of the yrast\cite{Kho96} and excited (band 3)
SD bands\cite{Hac96}.
The experimental Routhians of band 3 relative to the yrast SD band
have been extracted from these experimental data\cite{Hac96}.
The experiments indicate that band 3 is very low-lying,
$E'_x \approx 0.8$ MeV at $\w=0$,
which seems to support our interpretation of collective vibrations
because the lowest \tqp\ states have been predicted
to be at $E'_x \approx 1.5$ MeV in NMMS.

In Fig.~\ref{one} we compare the experimental Routhians
with the theoretical results presented in NMMS and Ref.\cite{Nak96}
(indicated by (1) and (2), respectively)
and with Routhians calculated with slightly different parameters
(indicated by (3)).
The parameter sets used for the calculations are;
(1) the same parameters as in NMMS 
(dynamically reduced pairing and $f_3=1$),
(2) the same as in Ref.\cite{Nak96}
(constant pairing gaps and $f_3=1.05$),
and (3) the same as (1) except $f_3=1.057$ and
$\hbar\omega_{\rm c}=0.5$ MeV for protons (see eq.(3.4) in NMMS).
It turns out that the previous calculation (2) had predicted
the experimental Routhians very nicely,
while the calculation (1) had overestimated the excitation energy by
about $200\sim 300$ keV.
The parameters (3) are chosen to reproduce the experiments:
it seems to suggest that the reduction of proton pairing is smaller than
expected in NMMS and the optimal octupole coupling strengths are
slightly larger than the harmonic-oscillator values.
This is consistent with the discussion in NMMS in which we have shown
the slightly larger coupling strengths ($f_3=1.05$) reproduce
the experimental $\J$ even better.

Experimental information has also been obtained on
the signature splitting of
bands 2 and 3 (assuming they are signature partners).
This is discussed in a paper by F.~Stephens
in this proceedings\cite{Ste96}.
Again, the agreement becomes better for the result with
the slower pairing reduction
and the larger octupole coupling.

\vspace*{-0.4cm}
\begin{minipage}{0.45\textwidth}
\begin{figure}[bh]
\epsfxsize=0.67\textwidth
\centerline{\epsfbox{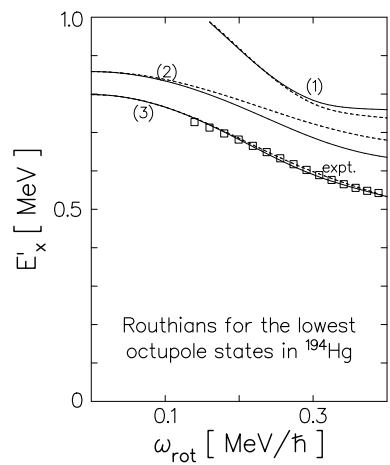}}
\vspace{-0.3cm}
\caption{
The RPA eigenenergies for the lowest octupole states
in \hgf.
Solid (dashed) lines correspond to states with signature
$\alpha=1$ (0).
Experimental data for band 3 ($\alpha=1$) are shown by
open squares.
}
\label{one}
\end{figure}
\end{minipage}
\newpage
\end{multicols}
 
\vspace*{-0.7cm}
\begin{figure}[th]
\epsfxsize=\textwidth
\centerline{\epsfbox{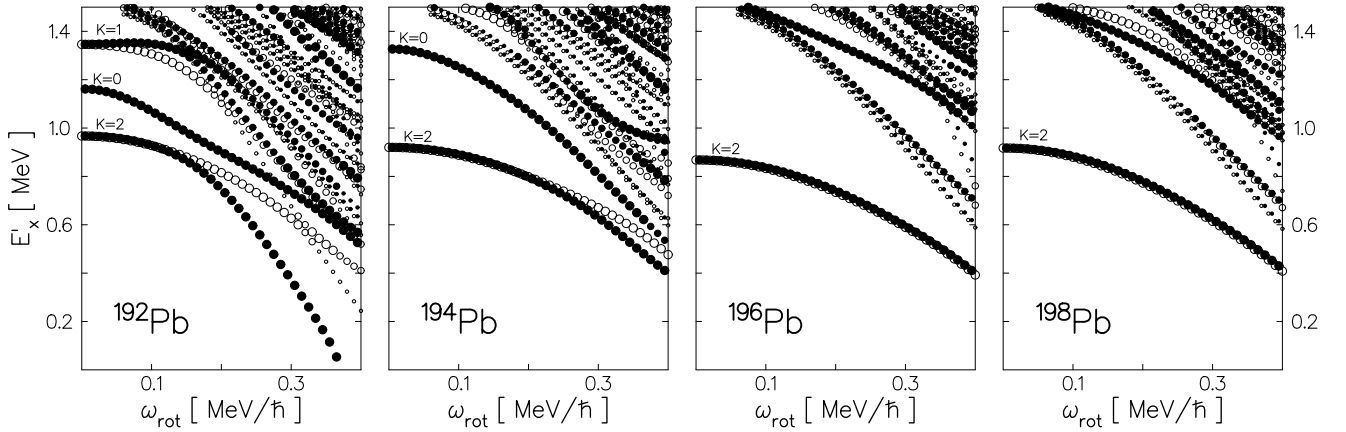}}
\caption{
RPA eigenenergies of negative-parity states for
$^{192,194,196,198}$Pb calculated with $f_3=1.05$.
Open (solid) circles indicate states with $\alpha=0$ (1).
Large, medium, and small circles indicate RPA solutions with $E3$
transition amplitudes larger than 200 $e\mbox{fm}^3$,
larger than 100 $e\mbox{fm}^3$, and less than 100 $e\mbox{fm}^3$,
respectively.
If we take $f_3=1$, the excitation energy of
the lowest octupole states will be around 1.2 MeV at $\w=0$.
}
\label{two}
\end{figure}

\ 
\vspace*{-0.7cm}
\begin{multicols}{2}
\section{Collective excitations in P\lowercase{b} nuclei}
\label{sec:pb}
In this section, the results are presented for excited states in
SD even-even Pb nuclei.
In $^{192,194,196,198}$Pb,
the predicted properties of $\gamma$ vibrations are similar
to those in Hg nuclei (see the previous section and NMMS);
the excitation energy ($E_x\approx 1.5\sim 1.6$ MeV)
is higher than the lowest octupole state
and their weak collectivity will be dissipated at high spin.
Therefore, the observation of the $\gamma$ vibration is expected to be
more difficult than the octupole vibrations.
Hereafter, let us focus our discussion on
the collective octupole excitations.

Figure~\ref{two} shows the calculated Routhians with negative parity
relative to the yrast SD bands for even-even Pb nuclei.
The quadrupole deformation $\ep=0.44$, the pairing parameters
$\Delta(0)=0.8$ (0.7) MeV with $\hbar\omega_c=0.5$ (0.5) MeV for
neutrons (protons), and $f_3=1.05$ are used throughout
(cf. sec.III-A in NMMS).
The lowest states are again the $K=2$ octupole vibrations for
all nuclei.

\vspace{-0.4cm}

\subsection{$^{194}$Pb and $^{196}$Pb}

\vspace{-0.3cm}

The excited SD bands in even-even Pb nuclei have been observed in
\Pbf\ (bands 2 and 3)\cite{Hug94} and \Pbs\ 
(bands 2, 3, and 4)\cite{Fai96}.
We assume the lowest octupole bands ($\alpha=0$ and 1) correspond
to the observed excited bands (bands 2 and 3) in \Pbf\ and \Pbs.
Unfortunately
we could not give unique assignment to band 4 in \Pbs\ (see below).

The calculated $\J$ moments of inertia are shown
in Fig.~\ref{three}(a)
together with the experimental data.
The signature of excited bands is determined by following the
experimental suggestions\cite{Hug94,Fai96}.
As is discussed in sec.\ref{sec:theory},
the $\J$ identical to those of the yrast SD band are
reproduced by means of the ``smearing'' effect.
The results calculated with $f_3=1.05$ agree with the
experiments especially well.
There is an experimental suggestion that\break

\vspace*{-1.7cm}
\begin{minipage}{0.45\textwidth}
\begin{figure}[th]
\epsfxsize=\textwidth
\centerline{\epsfbox{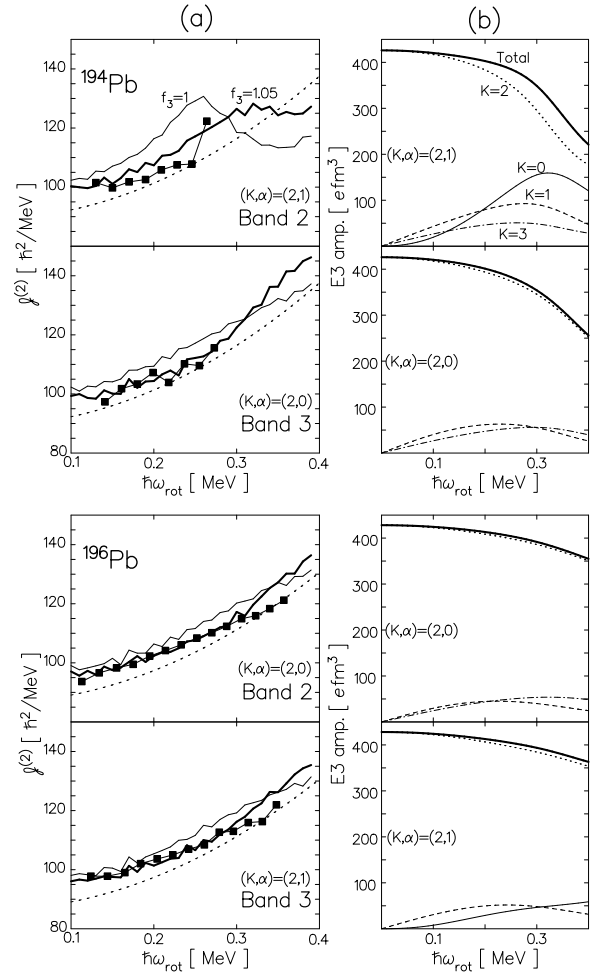}}
\caption{
(a) Calculated (solid lines) and experimental (symbols)
$\J$ moments of inertia for excited SD bands in \Pbf\ (upper)
and \Pbs\ (lower).
Thin solid lines are the results with $f_3=1$
while the thick lines indicate the results with $f_3=1.05$.
Dotted lines indicate the yrast $\J$.
Experimental data are taken
from Ref.\protect\cite{Hug94,Fai96}.\protect\\
(b) $E3$ transition amplitudes of the corresponding octupole states
calculated with $f_3=1.05$.
}
\label{three}
\end{figure}
\end{minipage}

\noindent
the band-head energy of
bands 2 and 3 in \Pbs\ is around 870 keV\cite{Fai96},
which again indicates better agreement with the results with $f_3=1.05$
(Fig.~\ref{two}).

In order to clarify how the ``smearing'' works,
we show examples of RPA amplitudes calculated with
$f_3=1.05$ for the lowest
$\alpha=1$ octupole state in \Pbs:
At $\w=0$,
the largest component is a two-quasineutron $\nu(56)$
(occupying $N=5$ and $N=6$ orbitals)
with $|\psi|^2=0.14$,
while it will be a two-quasiproton $\pi(56)$ with $|\psi|^2=0.12$
at $\hbar\w=0.3$ MeV.
As you can see, even the largest component occupies less than
10\% of the total sum of forward amplitudes
$\sum_{\mu\nu} |\psi(\mu\nu)|^2\approx 1.48$.
As is discussed before,
the neutron $N=7$ orbitals give the most significant effect on
$\J$ moments of inertia.
The largest component associated with neutron $N=7$ is
$|\psi|^2=0.04$ (0.02) at $\hbar\w=0$ (0.3) MeV.
Therefore, the blocking of $N=7$ orbitals turns out to be
extremely weak in this collective excitation,
which results in the $\J$ almost identical to those of the yrast
band.

The experiment\cite{Fai96} has also observed decay transitions
from excited SD bands (bands 2, 3, and 4)
into the yrast SD band in \Pbs\ 
(the dipole character of decays from band 3 was confirmed).
Assuming the $E1$ transitions,
the experimental B($E1$) values have been extracted from
the branching ratio;
B($E1)_{\rm exp}\approx 10^{-6}$, $10^{-5}$, and $10^{-4}$ W.u.
for bands 2, 3, and 4, respectively.
Using the $E1$ recoil charge ($-Ze/A$ for neutrons and $Ne/A$
for protons),
the calculations with $f_3=1.05$ at $\hbar\w=0.3$ MeV have suggested
B($E1)_{\rm cal}\approx 10^{-7}$, $10^{-6}$ W.u.
for bands 2 and 3, respectively.
If we assume band 4 is an either $K=0$ or 1 octupole band,
the B($E1)_{\rm cal}$ would be about $10^{-5}$ W.u.
Although the calculation underestimates the absolute magnitude
by a factor of 10,
the relative difference among bands 2, 3, and 4 is well reproduced.
However, the band-head energies of $K=0$ and 1 octupole states
are predicted to be around 1.5 MeV which is much higher than
the experimental suggestion ($\approx 1$ MeV);
this weakens our interpretation of band 4 as an octupole state.
Note that,
since we could not make the reliable prediction about $\beta$ vibrations,
we cannot deny a possibility that band 4 is a $\beta$ vibration.
Besides, if the transitions from band 4 are $M1$, it could be a \tqp\ band,
because the $\J$ moments of this band show the atypical behavior
which might suggest effects of the neutron $N=7$ orbitals.

Figure~\ref{three}(b) shows the $E3$ amplitudes ($K=0$, 1, 2, and 3)
of the lowest octupole states as functions of
frequency.
The $K$-mixing turns out to be weak for both bands 2 and 3
in \Pbs.
Since the $E1$ strength is supposed to be carried by
the low-$K$ component ($K=0$ and 1),
it provides the small B($E1$) for bands 2 and 3
($10^{-7}\sim 10^{-6}$ W.u.).
The calculation suggests the relatively strong Coriolis
mixing for $(K,\alpha)=(2,1)$ octupole band in \Pbf.
This leads to B($E1)_{\rm cal}\approx 10^{-6}\sim 10^{-5}$ W.u.
which is the largest among the octupole states shown in
Fig.~\ref{three}.
Further experimental investigation about the linking transitions
between excited and yrast SD bands in \Pbf\ 
may clarify the octupole collectivity of this band.

\subsection{$^{192}$Pb and $^{198}$Pb}
Although the excited SD bands in even-even A=190 nuclei
have been observed so far
only in $^{190,192,194}$Hg and $^{194,196}$Pb,
we expect this will be significantly extended in near future
by means of the new generation $\gamma$-ray detectors.
Among those candidates,
\Pbt\ and \Pbe\ may be relatively easy to access because the yrast
SD bands have been already observed.
In this section, we make a prediction on the properties of
octupole bands in these nuclei.

The octupole states in \Pbt\ are similar to those
(bands 2 and 4) in \hgo;
the lowest octupole phonon with signature $\alpha=1$ is
rotationally aligned, and the second lowest with $\alpha=0$ is
crossed by a \tqp\ band (Fig.~\ref{two}).
In Fig.~\ref{four}, we show the calculated $\J$ moments of inertia
for the lowest octupole bands in each signature sector.
The $\alpha=1$ band shows the large and almost constant $\J$,
while the $\alpha=0$ shows a bump at frequency
$\hbar\w \approx 0.3$ MeV (if we take $f_3=1$, the position of this
bump will be shifted to $\hbar\w \approx 0.23$ MeV).
Since the aligned octupole phonon is a result of strong Coriolis mixing,
the $\alpha=1$ band has substantial amounts of $K=0$ and 1 components
at finite frequency,
which will lead to the relatively strong $E1$
decays into the yrast SD band
(B($E1)_{\rm cal}\approx 10^{-6}\sim 10^{-4}$ W.u.).

In \Pbe, the lowest $K=2$ octupole states ($\alpha=0$ and 1)
are predicted to have no signature splitting.
Their $\J$ moments of inertia are identical to each other and similar
to those of the yrast SD band (Fig.~\ref{four}).
In this case, since the Coriolis mixing is very weak,
the $E1$ strengths are predicted to be small
(B($E1)_{\rm cal}\approx 10^{-7}\sim 10^{-6}$ W.u.).

\begin{minipage}{0.45\textwidth}
\begin{figure}[th]
\epsfxsize=1.05\textwidth
\centerline{\epsfbox{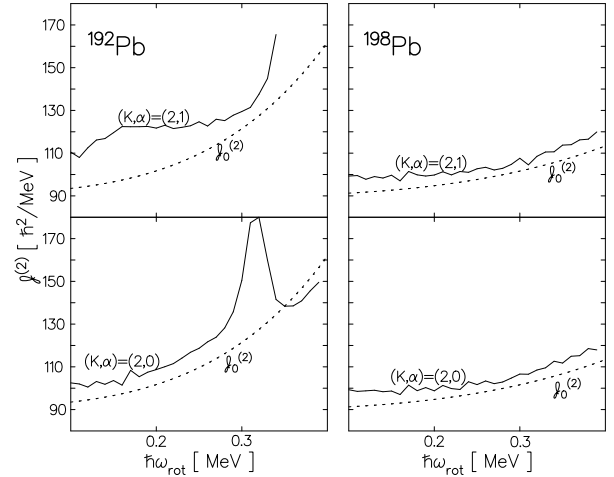}}
\caption{
Calculated $\J$ moments of inertia for the lowest octupole
bands with $\alpha=0$ (lower) and $\alpha=1$ (upper)
in \Pbt\ (left) and \Pbe\ (right).
$f_3=1.05$ is used in the calculations.
Dotted lines indicate the yrast $\J$.
Experimental data are taken from Ref.\protect\cite{Hen91,Cla94}.
}
\label{four}
\end{figure}
\end{minipage}

\newpage

\section{Conclusions: ``octupole paradise''}
The microscopic structure of the $\gamma$ and the octupole vibrations
built on the SD yrast bands in $^{190,192,194}$Hg and
$^{192,194,196,198}$Pb were investigated by means of
the RPA based on the cranked shell model.
For all these even-even nuclei, the $K=2$ octupole vibrations are
predicted to be the lowest.

New experimental data on SD \hgf\cite{Hac96}
seem to support our interpretation in NMMS 
that bands 2 and 3 may be $K=2$ octupole vibrational states.
The results in Ref.\cite{Nak96}
calculated with slightly different parameters
agreed better with the experimental Routhians for band 3,
which may suggest
the pairing reduction at finite frequency should be smaller
and the optimal octupole coupling strengths be larger than
the ones used in NMMS.

For the observed excited bands in SD Pb isotopes,
the following configurations have been assigned:\\
\begin{flushleft}
\begin{tabular}{lll}
\Pbf & Band 2 :&
the $(K,\alpha)=(2,1)$ octupole vibration.\\
     & Band 3 :&
the $(K,\alpha)=(2,0)$ octupole vibration.\\
\Pbs & Band 2 :&
the $(K,\alpha)=(2,0)$ octupole vibration.\\
     & Band 3 :&
the $(K,\alpha)=(2,1)$ octupole vibration.\\
     & Band 4 :&
indefinite.\\
\end{tabular}
\end{flushleft}

\noindent
With these assignments, our calculation accounted for
the $\J$ moments of inertia and
the observed decays of excited bands into the yrast SD band
(\Pbs).
It is also suggested that the relatively strong
Coriolis mixing
in the $(K,\alpha)=(2,1)$ octupole vibration in \Pbf\ 
may lead to the strong $E1$ decay into the yrast SD band.
It would be interesting for the experiment
to search for these decay transitions in this nucleus.

We have also done the calculations on excited SD bands in
\Pbt\ and \Pbe.
The following octupole bands are predicted
(possible experimental signatures are indicated in [{\it italics} ]):\\
\vspace{-0.5cm}
\begin{flushleft}
\begin{tabular}{lll}
\Pbt & (i) &
the $\alpha=1$ aligned octupole vibration\\
     &     & [{\it large $\J$ and $E1$ linking transitions}].\\
     & (ii) & the $(K,\alpha)=(2,0)$ octupole vibration crossed\\
     &     & by a \tqp\ band\\
     &     & [{\it a bump of $\J$ at $\hbar\w=0.2\sim 0.3$ MeV}].\\
\Pbe &\multicolumn{2}{l}{\ \ the signature-paired $K=2$ octupole vibrations}\\
     &\multicolumn{2}{l}{\ \ [{\it $\J$ similar to those of the yrast}].}
\end{tabular}
\end{flushleft}

As mentioned above,
the $K=2$ octupole vibrations are predicted to be the lowest
in the region where the SD bands have been observed so far
($79 \leq Z \leq 83$, $109 \leq N \leq 116$).
Since the $E1$ strengths only come from the Coriolis mixing
of the $K=0$ and 1 components,
the most direct evidence of octupole correlations,
namely the strong decays into the yrast SD band,
have been observed in limited cases
(band 2 in \hgo\ and band 4 in \Pbs).
However, if future experiments extend this region into
$N<108$ or $N>116$,
the $K=1$ octupole vibrations are predicted to become the lowest
(or close to the lowest).
Then, strong $E1$ decay should be observed,
and would be good experimental evidence 
for octupole collectivity.
This lowering of $K=1$ octupole states at open shell configurations
is a result of the striking shell structure at superdeformation
and has been discussed in Ref.\cite{MSM91,NMM92,Miz93}.

From these calculations and from a comparison with available
experiments,
we would like to conclude that the octupole vibrations are
prevalently observed in even-even SD A=190 nuclei
(``{\it Octupole Paradise}'').
The observation of the high-spin collective excitations
is very difficult in ND nuclei,
because the energy gaps at the Fermi surface quickly disappear
due to many aligned \tqp\ states,
so that non-collective modes of excitation become dominant at high spin.
The large deformation and large shell gap in SD nuclei
may overcome this situation
and provide us with a valuable opportunity
to observe a variety of collective excitations
in rapidly rotating quantum systems.

We would like to acknowledge F.~Azaiez, P.~Fallon,
G.~Hackman, M.A.~Riley, M.-G.~Porquet, F.~Stephens
for offering us new experimental data
and for valuable discussions.


\end{multicols}
\end{document}